\documentclass[12pt,preprint]{aastex}
\usepackage{psfig,rotate}

\shortauthors{Zeh, Klose, \& Hartmann}
\shorttitle{GRB supernovae}

\def\gr{\hbox{ \raisebox{-1.0mm}{$\stackrel{>}{\sim}$} }}
\def\kr{\hbox{ \raisebox{-1.0mm}{$\stackrel{<}{\sim}$} }}

\epsscale{0.6}

\begin{document}

\title{A Systematic Analysis of Supernova Light in Gamma-Ray Burst Afterglows}

\author{A. Zeh, S. Klose}
\affil{Th\"uringer Landessternwarte Tautenburg, 07778 Tautenburg, Germany}
\author{D. H. Hartmann}
\affil{Department of Physics and Astronomy, Clemson University, Clemson, SC
       29634-0978}

\date{Revised \today / Accepted}

\begin{abstract}
We systematically reanalyzed all Gamma-Ray Burst (GRB) afterglow data
published through the end of 2002, in an attempt to detect the predicted
supernova light component and to gain statistical insight on its
phenomenological properties. We fit the observed photometric light curves as
the sum of an afterglow, an underlying host galaxy, and a supernova
component. The latter is modeled using published multi-color light curves of
SN 1998bw as a template. The total sample of afterglows with established
redshifts contains 21 bursts (GRB 970228 - GRB 021211). For nine of these GRBs
a weak supernova excess (scaled to SN 1998bw) was found, what makes this  to
one of the first samples of high-$z$ core collapse supernovae. Among this
sample are all bursts with redshifts less than $\sim$0.7. These results
strongly support the notion that in fact all afterglows of long-duration GRBs
contain light from an associated supernova. A statistics of the physical
parameters of these GRB-supernovae shows that SN 1998bw was at the bright end
of its class, while it was not special with respect to its light curve
shape. Finally, we have searched for a potential correlation of the supernova
luminosities with the properties of the corresponding  bursts and optical
afterglows, but we have not found such a relation.
\end{abstract}


\section{Introduction \label{Intro}}

Observational and theoretical evidence suggest that the majority of GRB
progenitors are stars at the endpoint in stellar evolution (e.g., Fryer,
Woosley, \& Hartmann 1999; Heger et al. 2003). Since the discovery of a
near-by Type Ic supernova (SN 1998bw) in the error circle of the X-ray
afterglow for GRB 980425 (Galama et al. 1998; Kulkarni et al. 1998), evidence
is accumulating that core collapse supernovae are physically related to
long-duration GRBs. The supernova picture is further supported by the
observation that all GRB hosts are star-forming, and in some cases even
star-bursting galaxies (e.g., Frail et al. 2002; Sokolov et
al. 2001). Evidence for host extinction by cosmic dust in GRB afterglows
(e.g., Castro-Tirado et al. 1999; Klose et al. 2000), and the discovery of an
ensemble  of optically 'dark bursts' (for a recent discussion, see Fynbo et
al. 2001; Klose et al. 2003; Lazzati, Covino, \& Ghisellini 2002) also  is
consistent with the picture that GRB progenitors are young, massive stars
(Groot et al. 1998; Paczy\'nski 1998). Furthermore, for several GRB afterglows
X-ray data suggest a period of nucleosynthesis preceding or accompanying the
burst (e.g.,  Antonelli et al. 2000; Lazzati, Campana, \& Ghisellini 1999;
M\'esz\'aros \& Rees 2001). The angular distribution of the afterglows with
respect to their hosts also favors a physical relation of young, massive stars
to GRBs (Bloom, Kulkarni, \& Djorgovski 2002).

As a natural consequence of a physical relation between the explosion of
massive stars and GRBs supernova light should contribute to the afterglow
flux, and even  dominate under favorable conditions. The most convincing
example is GRB 030329 (Peterson \& Price 2003) at $z$=0.1685 (Greiner et
al. 2003a) with spectral confirmation of supernova light in its afterglow
(Hjorth et al. 2003; Kawabata et al. 2003; Matheson et al. 2003; Stanek et
al. 2003). In contrast to this unique spectroscopic evidence, several cases of
photometric evidence for extra light in GRB afterglows have been reported,
starting with GRB 980326 (Bloom et al. 1999). Various groups successfully fit
SN 1998bw templates to explain these late-time bumps, the most convincing case
being that of GRB 011121 (Bloom et al. 2002; Garnavich et al. 2003; Greiner et
al. 2003b).

The goal of the present paper is to analyze the supernova bumps in GRB
afterglow light curves using a systematic approach. While this was done also
for several bursts by Dado, Dar, \& de R\'ujula (2002a; and references
therein) with respect to a verification of their cannonball model (Dar \& de
R\'ujula 2003), we tackle this issue in an independent and different
way. First, from the numerical site, we have developed our own computational
approach. This includes a numerical procedure to redshift SN 1998bw light
curves (see \S~\ref{SNfits})  and to fit afterglow data within the context of
the fireball model. Second, from the observational site, when necessary and
possible we have performed late-time observations of some GRB host galaxies
(\S~2). A considerable part of the data we  have included in our study is
based on observing runs where we have been involved. Additional data have been
collected from the literature and checked for photometric consistency. Nearly
two dozen afterglows  had sufficient data quality, and a known redshift, to
search for a late-time bump in the light curve (\S~3). Third, we concentrate
our attention on a statistical analysis of the phenomenological parameters for
this class of GRB-SNe (\S~4). In this respect  our investigation goes beyond
the approaches undertaken by others to  explain late-time bumps in individual
afterglow light curves.

\section{Observations and data processing}

Some of the GRB afterglows we analyzed had poorly sampled late-time data,
which made it difficult to find or determine the parameters of a SN bump (GRBs
000418, 991208, and 010921). In order to perform late-time photometry of these
GRB hosts, we carried out two observing runs at the Calar Alto 3.5-m telescope
on March 13--14 and May 23--25, 2003. Observations were performed using the
multi-purpose camera \it MOSCA, \rm which uses a SITe 2k$\,\times4$k CCD with
a plate scale of 0.32 arcsec per pixel. The field of view is $11\,\times\,11$
arcmin$^2$. During the first observing run the seeing varied between 1.4 and
1.6 arcsec; in May the seeing was better than 0.8 arcsec.  Data reduction
followed standard procedures.

Most of the light curves we investigated have been followed in more than one
photometric band. For each of the GRBs we chose the best-sampled light curve
as a reference light curve for the fit in the other photometric bands. In the
most cases this was the $R$ band light curve. In brief, our approach was as
follows. \it First, \rm we assumed that the afterglow slopes and break time
are the same in all filters (Eq.~\ref{AG}), in reasonable agreement with
observational data.  Consequently, once we fit the reference light curve of an
optical transient and deduced the corresponding afterglow parameters, we
treated them as fixed parameters when fitting the light curves of the optical
transient in other photometric bands. Thereby the fit procedure was based on a
$\chi^2$-minimization with a Levenberg - Marquardt iteration. \it Second, \rm
for the representation of the supernova component we used published $UBVRI$
data of the light curve of SN 1998bw (Galama et al. 1998) as a template, and
redshifted them to the corresponding cosmological distance of the burster
(\S~\ref{SNfits}). These light curves are different from band to band. In
addition,  we allowed a variation of the SN luminosity with respect to SN
1998bw and a stretch in time (Eq.~\ref{ot} in \S~\ref{numerics}). \it Third,
\rm  the host magnitude, which represents a constant component in the integral
light of the optical transient, was usually considered as a free
parameter. Only for GRB 011121 and 020405 we used host-subtracted magnitudes
to fit the light curves.

Before performing a numerical fit, the observational data was corrected for
Galactic extinction along the line of sight using the maps of Schlegel,
Finkbeiner, \& Davis (1998). This also holds for SN 1998bw, where we assumed
$E(B-V)$= 0.06 mag. We calculated the Galactic visual extinction according to
$A_V^{\rm Gal}$ = 3.1 $E(B-V)$, whereas the extinction in $U$ and $B$ were
obtained via Rieke \& Lebofsky (1985), and in $R_c$ and $I_c$  by means of the
numerical functions compiled by Reichart (2001).

\section{Results \label{results}}

Among the 36$\pm1$ GRBs with detected optical afterglows up to the end of
2002\footnote{http://www.mpe.mpg.de/$^{\sim}$jcg/grbgen.html}, 21 had
sufficient data quality, and a known redshift, for a meaningful search for an
underlying supernova component (Table~1).  Among them in nine cases evidence
for a late-time bump was found. The results are summarized in Table~2. A
general inspection of this table makes clear that the burst ensemble with
detected late-time bumps in their afterglows  separates into a group with
statistically significant evidence for a bump (990712, 991208, 011121, and
020405), mostly in at least two photometric bands, and a group with less
significant bumps (970228, 980703, 000911, 010921, and 021211). Given the fact
that evidence for these bumps has also been reported by other groups (with the
only exception being GRB 010921), we feel confident that the results
presented in Table~2 can be used for a first statistical approach to
understand this type of GRB-SNe.

The most interesting result is that our numerical procedure found evidence for
a late-time bump in \it all \rm GRB afterglows with a measured redshift
$z\kr$0.7.  We believe that the interpretation of these bumps as an underlying
supernova component is the most natural and observationally most founded
explanation.  Among the higher redshifted bursts, we confirm the finding by
Holland et al. (2001) of a possible bump in the afterglow of GRB 980703, the
discovery by Lazzati et al. (2001) of a bump in the afterglow of GRB 000911
($z$=1.06; Price et al. 2002b), and a bump in the afterglow of GRB 021211,
which was discovered by Della Valle et al. (2003) and is also discussed by
Dado, Dar, \& de R\'ujula (2003b).

For five afterglows (GRB 970508, 991216, 000418, 010222, 020813) with $0.7 <
z < 1.5$, we can only place upper limits on the luminosity of any underlying
supernova component. These five upper limits have typical uncertainties of a
factor of two. The only exception is GRB 020813 where the uncertainty is
much larger, so that no firm conclusions can be drawn here.
The remaining seven bursts in our sample (GRB 971214, 990123,
990510, 000301, 000926, 011211, 021004) have redshifts $z>1.5$ and, therefore,
have not been investigated here, since this would have required a substantial
extrapolation of the SN 1998bw template into the UV domain. Finally, the
late-time bump clearly  seen in the afterglow of GRB 980326 (Bloom et
al. 1999) is not included in our study, because the redshift of the burster is
not yet known.

As we have outlined in the previous section, we fit the SN component using the
light curves of SN 1998bw as a template. Thereby, we allowed the luminosity
and the light curve shape to be a free parameter. The former means a scaling
of the luminosity of SN 1998bw by a factor $k$ (Eq.~\ref{ot}), whereby $k$
always refers to the corresponding wavelength region in the redshifted SN
frame. Differences in the light curve shapes were modeled by means of a
stretch factor $s$, which allows the supernova light curve to develop slower
($s>1$) or faster ($s<1$) than the one of SN 1998bw (Eq.~\ref{ot}). In this
respect we follow Hjorth et al. (2003) to describe the light curve of GRB
030329/SN 2003dh. The advantage of such an approach is that we can use
these two parameters to explore the entire ensemble of GRB-SNe in a
statistical sense. Table~2 shows that for the bursts with the photometrically
best sampled late-time bumps in their optical light curves (GRB 011121,
020405) the deduced parameters $k$ and $s$ are consistent with each other in
different wavelength regions. A comparison of these parameters of the
nine afterglows with late-time bumps, which are at different redshifts, seems
to be a reasonable first approach in order to constrain the width of the
luminosity distribution of GRB-SNe.

\begin{figure*}
\plotone{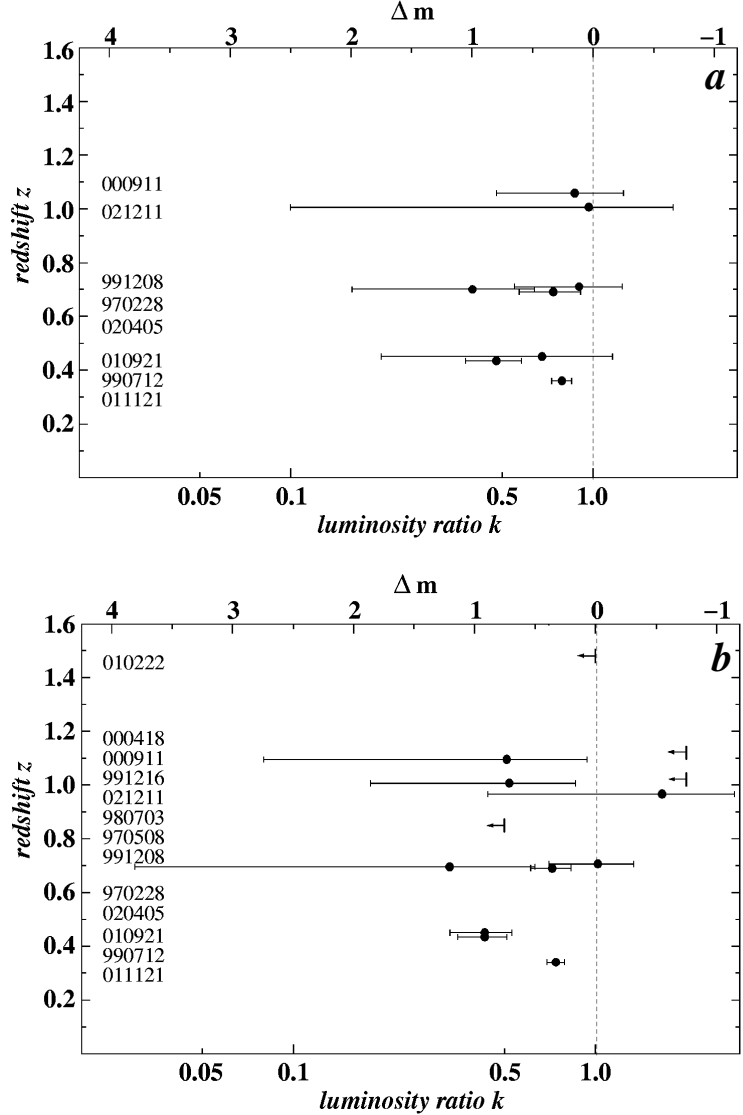}
\caption{The deduced  luminosities of GRB-SNe in units of
the luminosity of SN 1998bw in the same spectral region (parameter $k$,
Eq.~\ref{ot}). All data are based on observations in the $R$ band. The broken
line corresponds to SN1998bw; it is 
$\Delta m = -2.5$ log $k$, which  measures the
magnitude difference at maximum light between the GRB-SN and SN 1998bw in the
corresponding wavelength regime.  The lower panel (b)  is for a stretch
parameter fixed at $s$=1, while in the upper panel (a)  $s$ is not fixed. In
the former case we can set upper limits on $k$ for further four
bursts (GRB 970508, 991216, 000418, 010222).
Moreover, the numerical procedure can fit the afterglow light curve of
GRB 980703.  Note that the data are not corrected for a possible extinction in
the GRB host galaxies. \label{luminosity1}}
\end{figure*}

\begin{figure}
\plotone{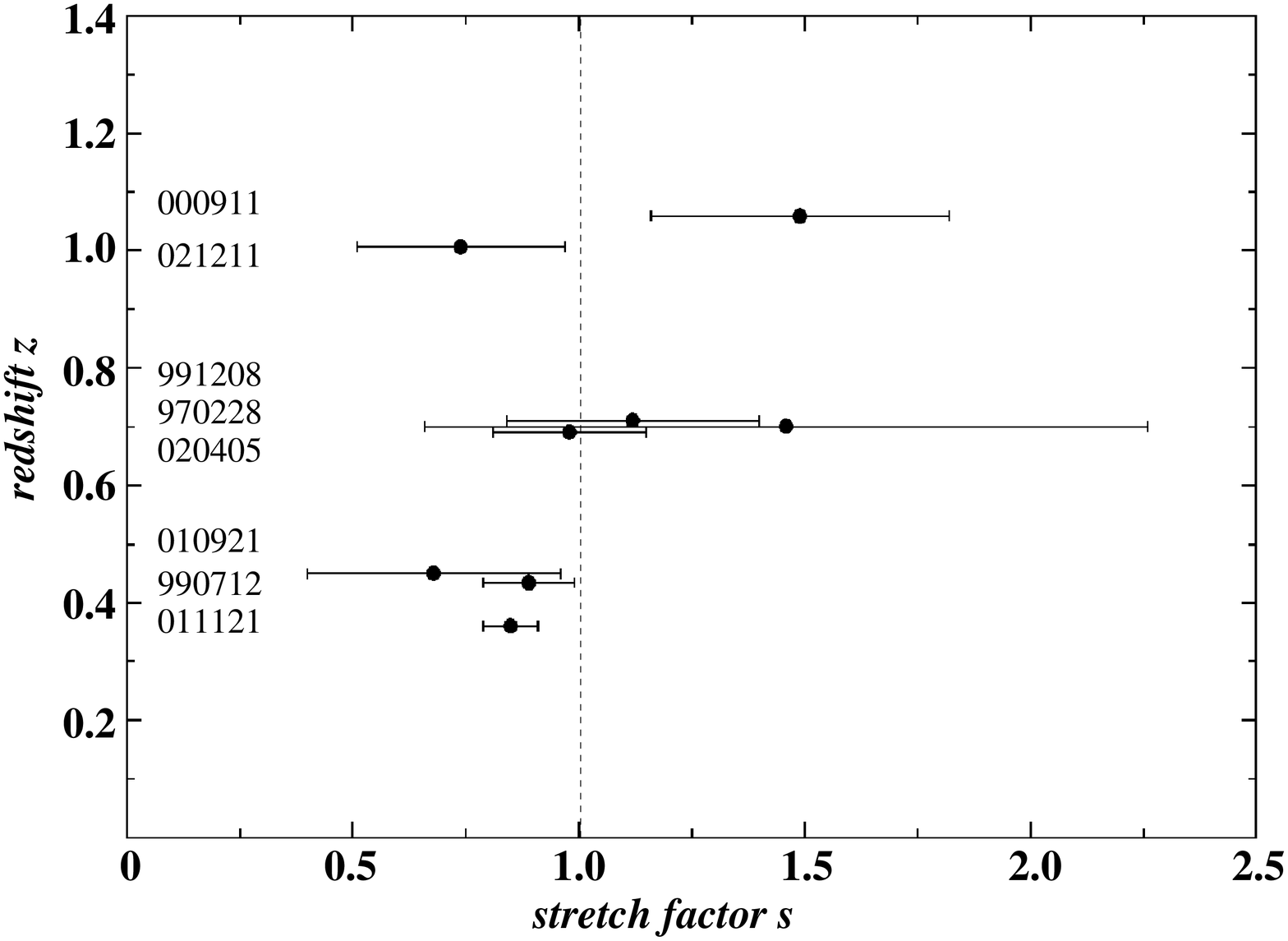}
\caption{The distribution of the
parameter $s$ (Eq.~\ref{ot}) describing a stretching of the SN light curve
relative to those of SN 1998bw (for which by definition $s=1$, broken line).
The mean value of $s$ is close to 1.0. \label{s}}
\end{figure}

\begin{figure}
\plotone{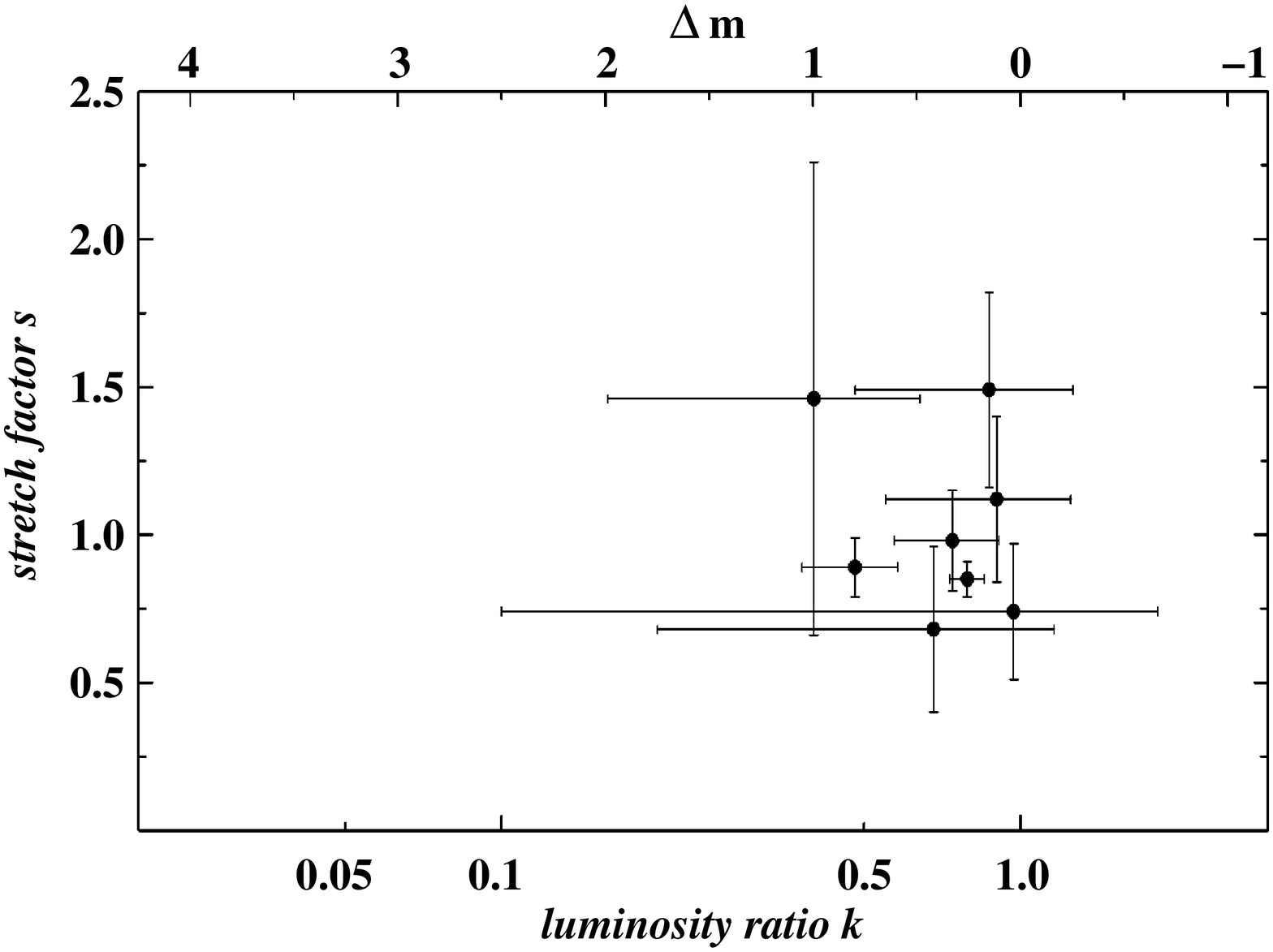}
\caption{Drawn here is the luminosity ratio $k$ versus
the stretch factor $s$ for the eight GRB-SNe of Figs.~\ref{luminosity1}a and
\ref{s}. No correlation between $k$ and $s$ is apparent in the
data. \label{sk}}
\end{figure}

In Fig.~\ref{luminosity1} we display the deduced parameter $k$ (luminosity)
for every individual GRB-SNe. We plot luminosity versus redshift just to look
for a potential evolutionary effect (which is not apparent) and to separate
the individual SNe from each other. While in  Fig.~\ref{luminosity1}a  we
allowed the stretch factor $s$ to be a free parameter during the numerical
fit, Fig.~\ref{luminosity1}b shows the results obtained when we fixed
$s=1$. The reason for the latter was twofold. First, if $s=1$ we can constrain
the luminosity of an underlying SN for those GRBs, where we do not detect a
bump in the late-time light curve. This is not possible in a reasonable way,
if we allow $s$ to be a free parameter. Second, sometimes the data base is too
small in order to include also the stretch parameter in the fitting procedure,
so that we have to fix $s$. Note that  in  Fig.~\ref{luminosity1} there are
three small sets of bursts at redshifts 0.4, 0.7 and 1.0. Between them is no
difference apparent neither in the luminosities of the GRB-SNe nor in the
width of the luminosity distribution. What is apparent from a comparison of
Fig.~\ref{luminosity1}a,b is that introducing the stretch factor reduces the
width of the luminosity distribution of the GRB-SNe and brings the luminosity
of all SNe a little closer to those of SN 1998bw ($k=1$). The distribution of
the deduced stretch factor itself is shown in Fig.~\ref{s}. Although $s$
varies by a factor of two  in both directions around $s=1$, within their
individual 1$\sigma$ error bars most data are  close to $s=1$. Finally,
no correlation was found between the deduced SN luminosity (parameter $k$)
and the stretch factor $s$ (Fig.~\ref{sk}).

\section{Discussion \label{statistics}}

\subsection{The luminosity distribution of the GRB-SNe}

\begin{figure}
\plotone{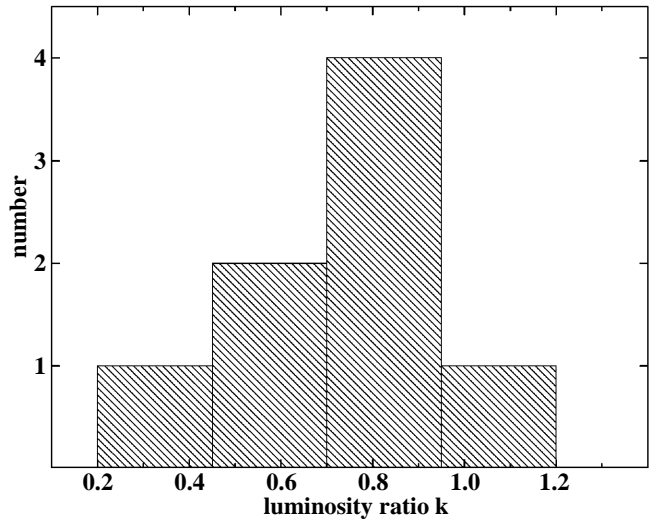}
\caption{The distribution
of the luminosity parameter $k$ (Eq.~\ref{ot}) as measured in the $R$ band in
the observer frame (Table~\ref{res}, with $s$ being a free parameter). GRB
980703 is not included here because the fitting procedure did  not find a
solution in this case. Note that the histogram does not include the
1$\sigma$ error bars of the individual $k$ factors, which are relatively
large. \label{histogram1}}
\end{figure}

\begin{figure}
\plotone{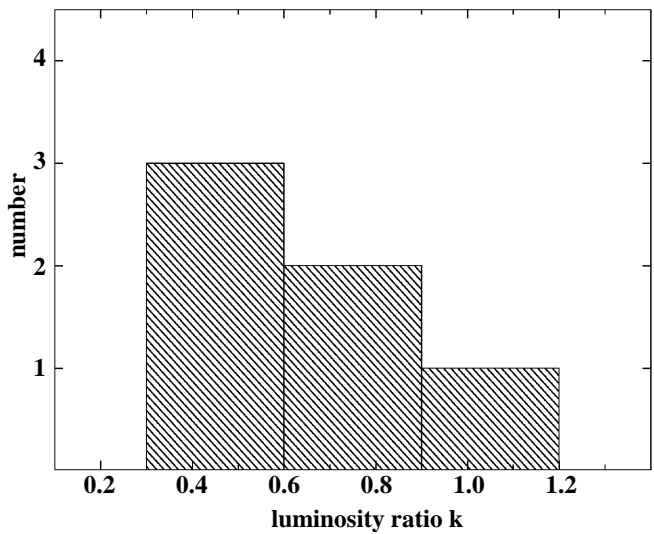}
\caption{The same as
Fig.~\ref{histogram1}, but for the  photometric band centered around
395$\pm$10 nm in the SN host frame (Table~\ref{res}, with $s$ being a free
parameter). Not included here are GRB 980703, 010921, and 021211 since there
is no data point in this wavelength range. \label{histogram2}}
\end{figure}

When plotting the parameter $k$ deduced for the $R$ band in the observer
frame, the width of the  luminosity  distribution of the class of GRB-SNe
(Fig.~\ref{histogram1}) is similar to what is observed for other classes of
core collapse SNe (Richardson et al. 2002). The mean of $k$ in the $R$ band is
0.7 independent of whether or not we fix the stretch parameter at $s=1$, while
our template SN 1998bw is at the bright end of the GRB-SN luminosity
distribution. The latter conclusion is supported when we plot $k$ for the same
wavelength region in the SN host frame, which is a better indicator of the
spread in SN luminosities. Most of the GRB-SNe we explored have a data point
for the photometric band centered around 395$\pm$10 nm in the corresponding
host frame (Table~\ref{res}). The distribution of $k$ now indicates again that
SN 1998bw is among the most luminous GRB-SNe. It also indicates that in fact
there is no peak around $k$=0.8 (Fig.~\ref{histogram1}), but we may so far
have only sampled the bright part of the GRB-SN luminosity function. Some
caution is of course required, given the partly large error bars of the $k$
factors, which are not shown in Figs.~\ref{histogram1} and
~\ref{histogram2}. While the conclusion that SN 1998bw was among the most
luminous members of its class seems to be robust, the shape of the GRB-SNe
luminosity function is still less well-determined. Extinction by interstellar
dust in the host galaxies could in principle also affect these results,
although only for GRB 010921 was a significant host extinction (\gr 1 mag)
reported (Price et al. 2003).

As we have outlined before, for redshifts $z$\kr0.7 all GRB
afterglows show evidence for an underlying late-time bump. Within
our context this means that we trace a complete set of GRB-SNe,
i.e., not only the brightest members of this class. The width of
this GRB-SNe luminosity distribution in the photometric band
centered around 395$\pm$10 nm in the SN frame is $\sim$1 to 1.5
magnitudes. This wavelength region is roughly placed in the $B$
band, so that we can compare the corresponding  luminosities with
those of other Type Ib/c supernovae (Richardson et al. 2002),
i.e., those class of SNe, which is believed to include the
progenitors of GRBs. It turns out that the GRB-SNe do fit into a
region between approximately $M_B=-19.5$ and $M_B=-18$ in figure 6
of Richardson et al., where no data on Type Ib/c SNe are known. If
all GRB-SNe are indeed of type Ib/c this would favor the
conclusion that the luminosity function of Type Ib/c SNe is rather
described by a broad  Gaussian than by a bimodal distribution
(figures 6 and 7 in Richardson et al.).

The non-detection of a supernova bump in more than half of the investigated
GRB afterglow light curves may be accounted for by  several reasons, like a
relatively bright host, or a faint supernova. In particular, finding a SN bump
for high-$z$ bursts is an observational challenge. For $z\gr0.7$  and $k=1$,
this peak magnitude exceeds $R_c=24$, which poses a major challenge for 3-m
class telescopes, given the usually very limited amount of target of
opportunity time for such observations. It is thus no surprise that only for
three of the GRBs above $z=0.7$ a supernova component was found (GRB 980703,
GRB 000911, GRB 021211), even though we can not rule out that the SN
'drop-out' is due to some evolutionary effect of the underlying burster
population and their environment.

\subsection{The SupraNova model}

Vietri \& Stella (2000) argued that GRBs are the result of delayed black hole
formation, which implies that the core collapse and its subsequent supernova
may significantly precede the burst. The delay could be of order months to
years (Vietri \& Stella 2000), or perhaps as short as hours (Woosley, Zhang, \&
Heger 2002). While constraining the latter possibility can not be accomplished
with the data at hand, the longer time scales are easily constrained. For only
two of the SN light curves the fit indeed improved if we allowed for a shift
in time between the onset of the burst and the onset of the SN (GRB 990712,
011121). The offsets never exceeded 5 days, and were both negative and
positive. However, the uncertainties in this parameter are large, due to the
poorly sampled shape of the underlying supernova (e.g., Garnavich et
al. 2003). The average shift is basically consistent with
zero. Presumably, these shifts are due to an underlying correlation between
luminosity and light curve shape, as observed in other types of supernovae
(e.g., Candia et al. 2003; Stritzinger et al. 2002). This is just what the
parameter $s$ takes into account. On the other hand, its clear that we have
no information about this issue in those cases where we have not found
evidence for a SN bump at late times. While this still leads open the
possibility of two populations of bursters (collapsars \& SupraNovae), we
emphasize again that we find a late time bump in all afterglows with
redshift $z<0.7$.

\subsection{X-ray lines and supernova bumps}

The identification of late-time bumps in afterglow light curves with SN light
would benefit from observations in  the X-ray band (for the cannonball model,
e.g., Dado, Dar, \& de R\'ujula 2003a). If the X-ray lines seen in some
afterglows (e.g., Reeves et al. 2002) have their origin in the circumburster
medium (e.g., Lazzati et al. 2001)  and not in the exploding star (e.g.,
M\'esz\'aros \& Rees 2001) this would be difficult to reconcile with the
interpretation of a late-time bump with an underlying SN
component. Unfortunately, the majority of bursts with high-resolution \it
XMM-Newton \rm or \it Chandra \rm spectroscopic X-ray follow-up observations
have no well-observed optical light curves. Among the afterglows with a
detected optical late-time bump listed in Table~2 only for GRB 020405 such
observations exist (Mirabal, Paerels, \& Halpern 2003); no evidence for X-ray
lines has been found there. \it BeppoSAX \rm observed the afterglow of GRB
970228 (Frontera et al.  1998) with comparable low energy resolution, and no
X-ray lines have been reported.  Among those low-$z$ bursts with well-defined
optical light curves and no evidence for a late-time bump in the data, only
for GRB 970508 \it BeppoSAX \rm X-ray follow-up observations have been
published (Piro et al. 1999). Evidence for an iron line was found.  Although
one might add to the list of bursts with well-observed late-time light curves
and additional spectral information in the X-ray band GRBs 990123,  000926,
010222, 011211, and 021004, the redshift of these bursts was \gr 1.5, making
it more or less hopeless to find an underlying SN component in the available
data base (an upper limit for GRB 010222 is reported in
Fig.~\ref{luminosity1}b; Henden et al. 2004, in preparation).  While it is
very interesting that neither for GRB 020405 (Mirabal et al. 2003) nor for GRB
030329 (Tiengo et al. 2003) lines have been found in X-ray spectra of their
afterglows, at present we cannot confirm a possible anti-correlation between
the occurrence of X-ray lines and the appearance of SN light in GRB
afterglows. This important issue remains to be investigated in the \it Swift
\rm era.

\begin{figure}
\plotone{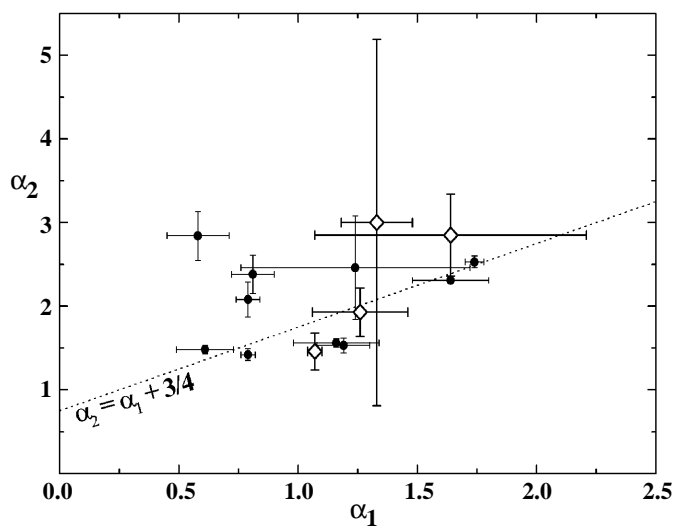}
\caption{The correlation
between the afterglow parameters $\alpha_1$ and $\alpha_2$ (Eq.~\ref{AG}) for
all afterglows with a break in their light curves. The dotted line is the
theoretical prediction in the simplest model ($\Theta_{\rm jet}$=const;
M\'esz\'aros \& Rees 1999). Open diamonds indicate the four
afterglows with a break and a detected underlying late-time bump, i.e., a SN
component (GRB 980703, 011121, 020405, 021211). 
Note that several afterglows listed in Table~\ref{allgrbs} 
showed no evidence for a break in their light curves, so
that they are not included in this figure. \label{a2a1e}}
\end{figure}

\subsection{SN properties vs. afterglow parameters}

Of particular interest is whether the burst and afterglow properties are to
some degree related to the existence of an underlying SN component.  For this
reason we have investigated if the deduced SN luminosity is correlated with
the corresponding energy release in the gamma-ray band (as given in Bloom,
Frail, \& Kulkarni 2003). No such correlation was found. We have also checked
whether the afterglow parameters $\alpha_1$ (pre-break decay slope),
$\alpha_2$ (post-break decay slope), and $t_b$ (break time; Eq.~\ref{AG}) from
those GRBs with detected SN component are different from  those without such a
component. Again, no correlation was apparent, even though one should keep in
mind that the data base is still very small. For illustration, in
Fig.~\ref{a2a1e} we display the relation between the afterglow parameters
$\alpha_1$ and $\alpha_2$ for all GRB afterglows we have investigated. While
there is a 'forbidden region' with $\alpha_2 \kr \alpha_1 + 3/4$ apparent in
the data, with the border line representing the simplest jet model with no
sideways expansion of the jet (M\'esz\'aros \& Rees 1999), no bimodality in
this distribution is visible. A tendency in our data that GRB afterglows with
detected underlying SN component seem to prefer pre-break slopes $\alpha_1>1$
should not be overinterpreted, since the $\alpha_1<1$ sample includes several
bursts with  redshifts $z>1.2$, where the photometric  detection of an
underlying SN component is difficult. On the other hand, at least one
selection effect does occur here. If a bright SN component is apparent in the
data then the parameter $\alpha_2$ of the afterglow light curve is usually
much more difficult to quantify, because the late-time evolution of the
genuine afterglow is less well-defined. This problem is well seen in the
afterglow light curve of GRB 011121 (e.g., Greiner et al. 2003) and GRB 030329
(e.g., Lipkin et al. 2004).

\section{Summary}

In an attempt to study the GRB-SN association, we have re-analyzed in a
systematic way all GRB afterglow data published by the end of 2002. We have
found that in nine cases evidence for extra light at late times is apparent in
the optical afterglow light curves. In most cases this is seen in more than
one photometric band. This extra light can be modeled well as supernova
light peaking $(1+z) 15...20$ days after a burst. Our main finding is that all
GRB afterglows with redshift $z\kr0.7$ showed evidence for extra light at
later times. For larger redshifts the data base is usually not of sufficient
quality, or the SN is simply too faint, in order to search for such a feature
in the late-time afterglow light curve.

The cut off date of our sample (end of 2002) was chosen to ensure that all
GRBs had published follow-up observations. Since that date, five new
afterglows with redshifts have been established. All but one were at redshifts
above 0.7, and again for the only nearby event (GRB 030329) a supernova
component was established. This is consistent with the statistical inferences
from the sample of earlier long-duration GRBs and leads us to conclude that
the current world sample of GRB afterglow measurements provides strong
statistical support for the link between (long-duration) GRBs and the final
stages of massive star evolution. This conclusion basically agrees with
earlier reports by Dado, Dar, \& de R\'ujula (2002a; and references therein),
and is essentially independent of the underlying GRB model.  While so far only
one event (GRB 030329) allowed a direct spectroscopic confirmation of this
link, the larger photometric sample discussed here supports this idea by
statistical means.

Based on our sample of nine GRB-supernovae we have performed a first
statistical approach to get insight on the characteristic luminosities of this
type of supernovae. We have found strong evidence that SN 1998bw  is at the
bright end of the GRB-SN luminosity distribution, with the latter matching
well into what is known so far about the luminosities of the brightest members
of other types of core collapse supernovae (Richardson et al. 2002). While
GRB-SNe are not standard candles, their peak luminosities are
comparable to those of Type Ia. In fact, within the context of the SN
interpretation of the late-time bumps in afterglows, our results demonstrate
once more that the first years in GRB research have already provided a first
sample of high-$z$ core collapse SNe up to a redshift of 1. This sample might
grow rapidly in the near future if indeed all long-duration GRBs tell us when
and where in the universe a massive star explodes.

Some caution is of course required. First, there is some bias
in the sample of bursts with detected optical afterglows. None of the bursts
with a detected SN bump was classified as an X-ray rich burst or an X-ray
flash, and for none of them were X-ray lines reported in the literature. In
other words, it is still possible that bursts with SN bumps do not belong to
these classes of events (but see Fynbo et al. 2004).  Second, for most of the
bursts discussed here evidence for a SN bump is based on a very small number
of data points around the SN peak time (say, between 10 to 40 days after the
burst), with the most critical cases being GRB 991208 and 010921. However,
we see no reason why we should disregard these events.

In their discovery paper, Klebesadel, Strong, \& Olson (1973)  noted that a
potential relation of GRBs to supernovae might still be an option to explain
this new  phenomenon. While the model they refer to (Colgate 1968) does not
describe what is today believed to be the underlying GRB mechanism,
historically it is nevertheless remarkable that the first paper ever about
GRBs might have given the right hint on the underlying source population,
followed by many years of trial and error.

\acknowledgements

S.K. and A.Z. acknowledge financial support by DFG grant KL 766/12-1 and from
the German Academic Exchange Service (DAAD) under grant No. D/0103745. D.H.H.
acknowledges support for this project under NSF grant INT-0128882. A.Z.
acknowledges the receipt of a scholarship from the
Friedrich-Schiller-University Jena, Germany. We thank Nicola Masetti and
Eliana Palazzi, Bologna, for providing host-subtracted photometric data on the
afterglow of GRB 020405. This work has profited from the GCN data base
maintained by Scott Barthelmy at NASA and the \it GRB big table \rm maintained
by Jochen Greiner, Garching.  We thank Kevin Lindsay (Clemson) for a careful
review of the manuscript. We thank the referee for critical comments, which
helped  to improve the paper.

\appendix

\section{Numerical approach}

\subsection{The light curve of the optical transient \label{numerics}}

We model the light curve of the optical transient (OT) following a GRB as a
composite of afterglow (AG) light, supernova (SN) light, and constant light
from the underlying host galaxy. The flux density, $F_\nu$, at a frequency
$\nu$ is then given by
\begin{equation}
 F_\nu^{\rm OT}(t) = F_\nu^{\rm AG}(t) + k \, F_\nu^{\rm SN}(t/s)
 + F_\nu^{\rm host}\,.
\label{ot}
\end{equation}
Here, the parameter $k$ describes the observed brightness ratio (in the host
frame) between the GRB-supernova, and the SN template (SN 1998bw) in the
considered photometric band (in the observer frame). We allowed $k$ to be
different in every photometric band, but within a band independent of
frequency. The parameter $s$ is a stretch factor with respect to the used
template. We have also explored  the consequences of a shift in time between
the onset of the burst and the onset of the supernova explosion, as implied by
certain theoretical models (Vietri \& Stella 2000). Then, in Eq.~(\ref{ot})
$F_\nu^{\rm SN}(t/s)$ was replaced by $F_\nu^{\rm SN}(t+\tau)$.  Here,
$\tau=0$ refers to GRB 980425/SN 1998bw (Iwamoto et al. 1998). If  $\tau<0$
the SN preceded the onset of the GRB.

We describe the afterglow light curve by a broken power-law (Beuermann et al.
1999; Rhoads \& Fruchter 2001),
\begin{equation}
   F_\nu^{\rm AG}(t) = \mbox{const}\
   [(t/t_b)^{\alpha_1\,n}+(t/t_b)^{\alpha_2\,n}]^{-1/n}\,,
\label{AG}
\end{equation}
with const=$2^{1/n} \, 10^{-0.4m(t_b)}.$ Here $t$ is the time after the burst
(in the observer frame),  $\alpha_1$  is the pre-break decay slope of the
afterglow light curve, $\alpha_2$ is the post-break decay slope, and $t_b$
is the break time. The parameter $n$ characterizes the sharpness of the break;
a larger $n$ implies a sharper break.
In most cases the parameter $n$ (Eq.~\ref{AG}) had to be fixed, otherwise
the iteration did not converge. The reason was the usually too small number of
data points around the break time. In these cases we set $n=10$, producing a
relatively sharp break in the light curve. However, this procedure did not
strongly affect the deduced supernova parameters $k$ and $s$ (Eq.~\ref{ot}).

In the observer frame the flux density of the time-dependent supernova light
is given by (cf. Dado et al. 2002a)
\begin{equation}
F_\nu^{\rm SN}(t) = \frac{1+z_{\rm SN}}{1+z_{\rm bw}}\ \frac{d_{L,{\rm
      bw}}^2} {d_{L,{\rm SN}}^2}\, F_{\rm bw} \Big(\nu\ \frac{1+z_{\rm
      SN}}{1+z_{\rm bw}}; t \ \frac{1+z_{\rm bw}} {1+z_{\rm SN}}
      \Big)\,.
\label{Dado}
\end{equation}
Here 'SN' stands  for the GRB supernova under consideration, and 'bw'
represents SN 1998bw ($z$=0.0085; Tinney, Stathakis, \& Cannon 1998). We
calculated the luminosity distance, $d_L$, assuming a flat universe with
matter density $\Omega_M =$ 0.3, cosmological constant $\Omega_\Lambda=$ 0.7,
and Hubble-constant $H_0=65$ km s$^{-1}$ Mpc$^{-1}$.

We always fitted photometric magnitudes. After manipulating
Eqs.~(\ref{ot}, \ref{AG}), the apparent magnitude of the OT in a given
photometric band is given as
\begin{equation}
m_{\rm OT}(t) = -2.5\,\log\{10^{-0.4\,m_c}[(t/t_b)^{\alpha_1\,n}  +
          (t/t_b)^{\alpha_2\,n}]^{-1/n} +
          k\,10^{-0.4\,m_{\rm SN}(t/s)} + 10^{-0.4\,m_{\rm host}}\}\,.
\label{mag1}
\end{equation}
Again, $t/s$ was replaced by $t+\tau$ when we allow for a delay between SN and
GRB.  Equation~(\ref{mag1}) has eight free parameters: $\alpha_1, \alpha_2, n,
t_b, k, s, m_{\rm host}$, and the constant $m_c$, which absorbs the
constant of Eq.~(\ref{AG}) and corresponds to the magnitude of the fitted
light curve for the case $n=\infty$ at the break time $t_b$. If there is no
break in the light curve, then Eq.~(\ref{mag1}) reduces to
\begin{equation}
m_{\rm OT}(t) = -2.5\,\log\{10^{-0.4\,m_1}t^{\alpha} +
          k\,10^{-0.4\,m_{\rm SN}(t/s)} + 10^{-0.4\,m_{\rm host}}\}\,,
\label{mag2}
\end{equation}
where $m_1$ is the brightness of the afterglow at $t=1$ day after the
burst (if $t$ is measured in days).

\subsection{Redshifting the SN 1998bw light curves \label{SNfits}}

\begin{figure}
\epsscale{1.0}
\plotone{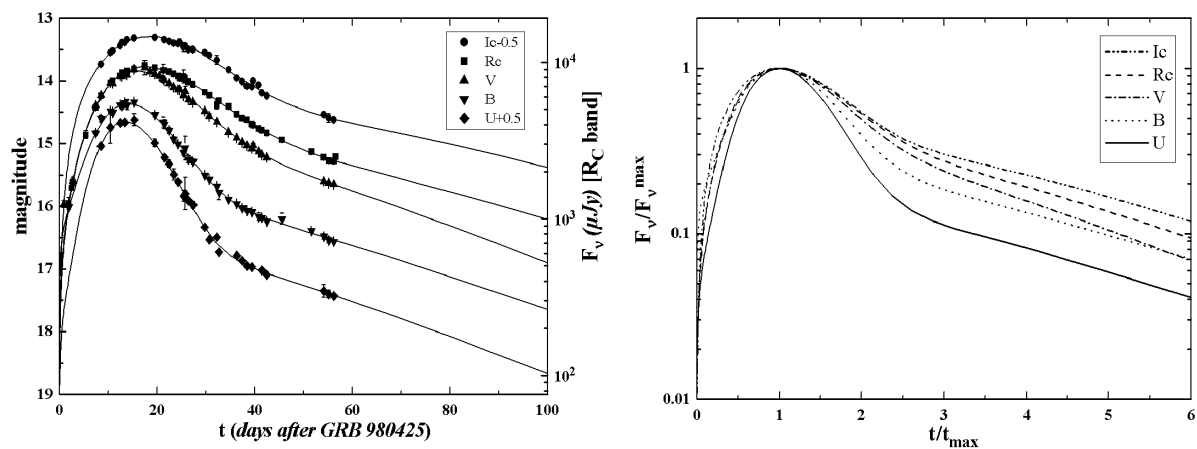}
\caption{\it Left: \rm The
$UBVRI$ light curves of SN 1998bw according to the data  provided by Galama et
al. (1998). The fits (drawn through lines) are based on a purely empirical
equation which fits a supernova light curve very well and is not physical. It
extrapolates also beyond 60 days. Note that the light curves differ in peak
flux, peak time, and in shape.  In order to predict the light curves of a
redshifted SN 1998bw one has to construct light curves for any frequency in
between the characteristic frequencies of the $UBVRI$ bands.  \it Right: \rm
The broad-band light curves of SN 1998bw normalized to
their peak maxima and peak times. Now they differ only in their shapes.
\label{98bw}}
\end{figure}

Equation~(\ref{ot}) requires as an input the function $F_\nu^{\rm SN}(t)$ for
arbitrarily frequencies in the optical bands. Spectra from SN 1998bw are
available in the literature, but the time coverage of published broad-band
photometry is much better. Therefore, we constructed $F_\nu^{\rm SN}(t)$ based
on published $UBVRI$ light curves (Galama et al. 1998), assuming that we can
smoothly interpolate between adjacent photometric bands. Thereby, we have
taken into account that various broad-band features are inherent to the
spectral energy distribution of SN 1998bw that develop with time (e.g., Patat
et al. 2001; Stathakis et al. 2000). Therefore, for different photometric
bands SN 1998bw light curves peak at different times $t_{\nu}^{\rm max}$, at
different flux densities, $F_\nu^{\rm max} = F_\nu(t_{\nu}^{\rm max})$, and
have different shapes.\footnote{For reasons of clarity, in this section we
omit the index 'SN' at $F_\nu$; all flux densities refer to SN 1998bw.} In the
following, we demonstrate our numerical approach using the $U$ and $B$ bands
as an example (Fig.~\ref{98bw}).

Let $\nu_U$ be the central frequency of the $U$ band, $\nu_B$ be the central
frequency of the $B$ band, and $\varepsilon$ be defined as $0 \le \varepsilon
\le 1$.  Let $\nu'$ be the frequency for which the light curve $F_\nu(t)$ is
required, then the relation $\nu' = \nu_U + \varepsilon \, (\nu_B - \nu_U)$
defines  the value of $\varepsilon$. Assuming a smooth behavior of $F_\nu (t)$
between the $U$ band and the $B$ band, for the frequency-dependent peak flux of
the light curve at the frequency $\nu'$, we assume
\begin{equation}
\log F_{\nu'}^{\rm max} = \log F_U^{\rm max} \nonumber \\ +
     \varepsilon \, (\log F_B^{\rm max} - \log F_U^{\rm max})\,.
\end{equation}
Similarly, for the frequency-dependent peak time of the supernova light curve
at a fixed frequency, $\nu'$, we write
\begin{equation}
t_{\nu'}^{\rm max} = t_U^{\rm max} + \varepsilon \, (t_B^{\rm max} -
                     t_U^{\rm max})\,.
\end{equation}

Finally, in order to model the frequency-dependent shape of the
SN 1998bw light curves, we normalize them to their peak flux and peak time.
Then, at a given frequency we have $F_\nu(t) = \eta_\nu \ F_{\nu}^{\rm max},$
where $\eta$ is a function of the ratio $t/t_\nu^{\rm max}$ and $0 \le \eta \le
1$. Correspondingly, our ansatz for the shape function $\eta_{\nu'}$ for a
redshifted SN 1998bw is
\begin{equation}
\log \eta_{\nu'}(x) = \log \eta_U (x) +  \varepsilon \,[\log \eta_B
      (x) - \log \eta_U(x) ]\,,
\end{equation}
where $x= t_{\rm host} / t_{\nu'}^{\rm max}$ and
\begin{equation}
t_{\rm host} = t \ \frac{1+z_{\rm bw}}{1+z_{\rm SN}}
\end{equation}
is measured in the host frame (symbols follow Eq.~\ref{Dado}).

Once an ensemble of functions $F_\nu (t)$ has been calculated, the apparent
magnitude of the redshifted SN 1998bw in a given photometric band
(Eqs.~\ref{mag1}, \ref{mag2}) is obtained by integrating over the flux density
per unit wavelength ($F_\lambda (t) = -\nu^2/c \, F_\nu (t)$), multiplied by
the corresponding filter response function $S_\lambda$. For $S_\lambda$ we
used the transmission curves for Bessel filters provided on the internet pages
of the \it European Southern Observatory \rm  for VLT-FORS1 with reference to
Bessel (1979). For the transformation between photometric magnitudes and flux
densities we used the calibration constants provided by Fukugita, Shimasaku,
\& Ichikawa (1995; their table 9) and Zombeck (1990). The so calculated
broad-band light curves of a redshifted SN 1998bw were then used as an input
function for Eqs.~(\ref{mag1}, \ref{mag2}).

The results of our numerical procedure were compared with corresponding
results published by Dado et al. (2002a) and Bloom et al. (2002), and we found
close agreement. We used our procedure to correctly \it predict \rm the color
evolution of GRB-SN 030329 (Zeh, Klose, \& Greiner 2003), and have performed a
very good numerical fit for the light curves of GRB-SN 011121 (Greiner et
al. 2003b). The limit of our procedure is given by the chosen photometric band
in combination with the redshift of the burster. Once we can no longer
interpolate in between the $UBVRI$ bands, but have to extrapolate into the UV
domain (cf. Bloom et al. 1999), results become less accurate.

\section{Notes on individual bursts with detected supernova bump
\label{individual}}

\it GRB 970228: \rm \ The light curves were constructed based on the data
compiled by Galama et al. (2000). The $R_c$ band light curve shows evidence
for extra light appearing $\sim$2 weeks after the burst, which can be
attributed to an underlying SN 1998bw component at the redshift of the burster
(Galama et al. 2000; Reichart 1999). There is no evidence for a break, but its
presence cannot be excluded due to the rather sparse data set. In the $I_c$
band the SN bump is most visible, but the light curve suffers from a
lack of early-time data. In the $V$ band the significance for extra light is
even lower, again due to the lack of observational data.

\it GRB 980703: \rm \ The search of an SN component in the afterglow of GRB
980703 is affected by the relatively bright host. Most data were taken
from Bloom et al. (1998), Castro-Tirado et al. (1999), Holland et al. (2001),
and Vreeswijk et al. (1999). Evidence for a late-time bump is rather weak.

\it GRB 990712: \rm \ This burst had a relatively bright host galaxy
hampering the long-term study of its afterglow. We used the data presented by
Fruchter et al. (2000a), Hjorth et al. (2000), and Sahu et al. (2000) to
analyze the light curves. Although the $R_c$ band light curve is well-sampled,
the bright host may have hidden the detectability of a break at later
times. Our numerical procedure finds evidence for extra-light, confirming the
finding by Bj\"ornsson et al. (2001).

\it GRB 991208: \rm \ We used the compilation of data by Castro-Tirado et
al. (2001), with additional data from Dodonov et al. (1999), Halpern et
al. (1999), Garnavich et al. (1999), and Fruchter et al. (2000b), including
our late-time observation of the host in early 2003 to analyze
the light curves. The $R_c$ band data can be fitted with or without the
inclusion of a break. In the former case the break is mainly due to a
single data point at $t\sim7$ days. Most likely, the afterglow was
discovered after a break had already occurred in the light curve. The
afterglow parameters were determined in the $R_c$ band. These parameters fit
well in the $V$ band. Unfortunately, in the $I_c$ band no data were obtained
during the time of the SN bump.

\it GRB 000911: \rm \ This was a long-lasting burst with a duration of
$\sim$500
seconds (Hurley et al. 2000; Price et al. 2002b). The optical afterglow
was in detail observed by Price et al. (2002b) and Lazzati et al. (2001).  We
confirm the finding by Lazzati et al. that the published data show evidence
for a bump in $VRI$ at later times.

\it GRB 010921: \rm \ This burst occurred in a rather crowded stellar field
and had a relatively large error box (Hurley et al. 2001), which hampered the
early detection of its afterglow (Price et al. 2001). The light curve is
therefore not well-sampled. Using the data published by Park et al. (2002) and
Price et al.  (2002a) combined with our late-time observations of the host,
our numerical procedure finds evidence for extra light with its peak time
$\sim2$ weeks after the burst. This result is obtained when we adopted a
single power-law decay, where the decay slope $\alpha$ was deduced from the
$r'$ band light curve. Our procedure finds a SN component with $k=0.68\pm0.48$
and $s=0.68\pm0.28$. Within the given uncertainties this is not in conflict
with the upper limit reported by Price et al. (2003).

\it GRB 011121: \rm \ This was the nearest known burst at the time of its
discovery (excluding GRB 980425/SN 1998bw). It showed clear evidence for an
underlying SN component in several photometric bands (Bloom et
al. 2002; Dado, Dar, \& de R\'ujula 2002b; Garnavich et al. 2003; Greiner et
al. 2003b). In our fit we included late-time \it Hubble Space Telescope \rm
data (Bloom et al. 2002). A break is apparent in the light curve at
$t\sim1$ day (see Greiner et al. 2003b; note that in Greiner et al. we
assumed a Galactic extinction towards SN 1998bw of 0 mag).

\it GRB 020405: \rm \ This is the second burst with known redshift and a
well-observed bump in its late-time afterglow. We used host-subtracted data
provided by N. Masetti (private communication) to analyze the light
curves. Extra light apparent in the late-time light curve can be attributed to
an underlying SN component, as already noted by Masetti et al. (2003).  Our
procedure also detects a break in the light curves at $t\sim2$ days.

\it GRB 021211: \rm \ For the fit we included data published by Della Valle et
al. (2003), Fox et al. (2003), Li et al. (2003), and Pandey et al. (2003). A
weak bump is apparent at late times, which is most likely due to an underlying
SN component given its (weak) spectral confirmation (Della Valle et
al. 2003).


\clearpage

\begin{table*}
\caption{The input sample of GRB afterglows\tablenotemark{a}\label{allgrbs}}
\vspace{0.3cm}
\begin{tabular}{|ll|ll|ll|}
\hline \noalign{\smallskip}
GRB & $z$ & GRB & $z$ & GRB & $z$ \\
\noalign{\smallskip} \hline
\noalign{\smallskip}
970228 & 0.695 & 991208 & 0.706 & 010921 & 0.450 \\
970508 & 0.835 & 991216 & 1.02  & 011121 & 0.362 \\
971214 & 3.42  & 000301C& 2.04  & 011211 & 2.140 \\
980703 & 0.966 & 000418 & 1.118 & 020405 & 0.69  \\
990123 & 1.600 & 000911 & 1.058 & 020813 & 1.25  \\
990510 & 1.619 & 000926 & 2.066 & 021004 & 2.3   \\
990712 & 0.434 & 010222 & 1.477 & 021211 & 1.01  \\
\noalign{\smallskip} \hline
\end{tabular}
\tablenotetext{a}{Redshifts were taken from the literature.}
\end{table*}

\clearpage

\begin{table*}
\caption{Best-fit parameters for the SN component found in GRB
afterglows\tablenotemark{a} \label{res}}
\renewcommand{\tabcolsep}{4pt}
\vspace{0.3cm}
\begin{tabular}{|lccc|ccc|cc|cc|}
\hline \noalign{\smallskip}
GRB  & $z$ & band & $\lambda_{\rm host}$& $k$ & $s$ & $\chi^2_{\rm d.o.f.}$
& $k$ if $s$=1& $\chi^2_{\rm d.o.f.}$ & $\chi^2_{\rm noSN}$ & data \\
\noalign{\smallskip} \hline
\noalign{\smallskip}
       &       & $I_c$ & 476 & \nodata       & \nodata       &\nodata& 0.66$\pm$0.27 & 0.01  & 2.16  & 4 \\
970228 & 0.695 & $R_c$ & 389 & 0.40$\pm$0.24 & 1.46$\pm$0.80 & 0.70  & 0.33$\pm$0.30 & 0.71  & 0.77  & 10\\
       &       & $V$   & 325 & \nodata       & \nodata       &\nodata& 0.25$\pm$0.50 & 0.06  & 0.15  & 4 \\[1mm]

       &       & $I_c$ & 410 & \nodata       & \nodata       &\nodata& \nodata       &\nodata& 1.59  & 14\\
980703 & 0.966 & $R_c$ & 335 & \nodata       & \nodata       &\nodata& 1.66$\pm$1.22 & 0.78  & 0.79  & 19\\
       &       & $V$   & 280 & \nodata       & \nodata       &\nodata& \nodata       &\nodata& 1.50  & 7 \\[1mm]

       &       & $I_c$ & 562 & 1.00$\pm$0.38 & 0.56$\pm$0.10 & 0.55  & \nodata       &\nodata& 1.67  & 6 \\
990712 & 0.434 & $R_c$ & 459 & 0.48$\pm$0.10 & 0.89$\pm$0.10 & 1.00  & 0.43$\pm$0.08 & 1.01  & 2.25  & 23\\
       &       & $V$   & 384 & 0.37$\pm$0.44 & 0.71$\pm$0.36 & 2.30  & 0.29$\pm$0.18 & 1.62  & 1.99  & 16\\[1mm]

       &       & $I_c$ & 472 & \nodata       & \nodata       &\nodata& \nodata       &\nodata& 1.82  & 13\\
991208 & 0.706 & $R_c$ & 386 & 0.90$\pm$0.35 & 1.12$\pm$0.28 & 1.64  & 1.02$\pm$0.32 & 1.56  & 2.52  & 20\\
       &       & $V$   & 323 & 1.16$\pm$0.19 & 1.86$\pm$0.10 & 0.45  & 0.93$\pm$0.30 & 1.04  & 2.26  & 11\\[1mm]

       &       & $I_c$ & 392 & 0.39$\pm$0.37 & 1.06$\pm$0.50 & 1.83  & 0.40$\pm$0.29 & 1.30  & 1.61  & 7 \\
000911 & 1.058 & $R_c$ & 320 & 0.87$\pm$0.39 & 1.49$\pm$0.33 & 0.75  & 0.51$\pm$0.43 & 1.14  & 1.22  & 8 \\
       &       & $V$   & 267 & \nodata       & \nodata       &\nodata& 0.43$\pm$1.24 & 1.25  & 1.42  & 6 \\[1mm]

       &       & $I_c$ & 556 & \nodata       & \nodata       &\nodata& 0.40$\pm$1.67 & 0.50  & 0.28  & 4 \\
010921 & 0.450 & $R_c$ & 454 & 0.68$\pm$0.48 & 0.68$\pm$0.28 & 0.42  & 0.43$\pm$0.10 & 0.78  & 2.74  & 6 \\
       &       & $V$   & 380 & \nodata       & \nodata       &\nodata& \nodata       &\nodata&\nodata& 2 \\[1mm]

       &       & $I_c$ & 632 & \nodata       & \nodata       &\nodata& \nodata       &\nodata&\nodata& 0 \\
011121 & 0.360 & $R_c$ & 484 & 0.79$\pm$0.06 & 0.85$\pm$0.06 & 0.92  & 0.74$\pm$0.05 & 1.32  & $>$20 & 13\\
       &       & $V  $ & 405 & 0.86$\pm$0.09 & 0.81$\pm$0.06 & 2.13  & 0.83$\pm$0.05 & 3.26  & $>$20 & 10\\[1mm]

       &       & $I_c$ & 476 & 0.76$\pm$0.17 & 0.80$\pm$0.17 & 5.29  & 0.71$\pm$0.10 & 5.68  & $>$20 & 10\\
020405 & 0.695 & $R_c$ & 389 & 0.74$\pm$0.17 & 0.98$\pm$0.17 & 5.26  & 0.72$\pm$0.11 & 4.86  & $>$20 & 18\\
       &       & $V$   & 325 & 0.69$\pm$0.22 & 0.74$\pm$0.13 & 6.79  & 0.53$\pm$0.16 & 6.91  & $>$20 & 14\\[1mm]

       &       & $I_c$ & 428 & \nodata       & \nodata       &\nodata& \nodata       &\nodata&\nodata& 0 \\
021211 & 1.006 & $R_c$ & 328 & 0.97$\pm$0.87 & 0.74$\pm$0.23 & 2.68  & 0.52$\pm$0.34 & 2.65  & 2.79  & 35\\
       &       & $V$   & 274 & \nodata       & \nodata       &\nodata& \nodata       &\nodata&\nodata& 0 \\
\noalign{\smallskip} \hline

\end{tabular}
\tablenotetext{a}{Columns: (1) and (2): GRB and redshift; (3) photometric
band, in which the light curve was fitted; (4) central wavelength of the
photometric band in the host frame in units of nm, adopting for $V, R_c, I_c$
wavelengths of 550, 659, and 806 nm, respectively; (5) peak luminosity of the
fitted SN component in the corresponding wavelength band (observer frame) in
units of SN 1998bw, after correction for Galactic extinction; (6) stretch
factor $s$ (Eq.~\ref{ot}); (7) goodness of fit per degree of freedom; (8) and
(9) the same as (5) and (6) for $s=1$; (10) goodness of fit per  degree of
freedom assuming that there is no underlying SN component; (11) total
number of data points used for the fit.  Note that the low
$\chi^2$/d.o.f. for GRB 970228 is due to the small number of data points.}
\end{table*}

\end{document}